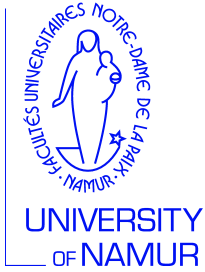

UNIVERSITY
OF NAMUR

NIMASTEP:
a software to modelize, study and analyze the dynamics
of various small objects orbiting specific bodies

by N. Delsate and A. Compère

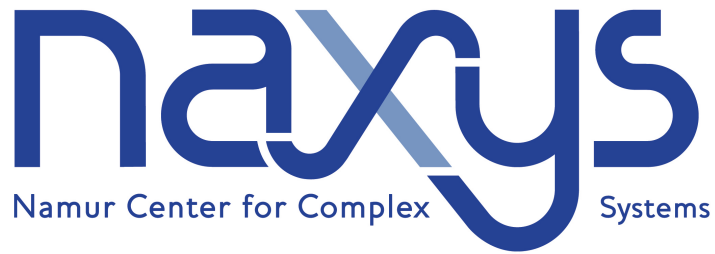

**Namur Center for Complex Systems**
University of Namur
8, rempart de la vierge, B5000 Namur (Belgium)

`http://www.naxys.be`

# NIMASTEP: a software to modelize, study, and analyze the dynamics of various small objects orbiting specific bodies

N. Delsate and A. Compère

University of Namur, NAXYS, Department of Mathematics, Rempart de la Vierge 8, B-5000 NAMUR,BELGIUM
e-mail: nicolas.delsate@math.fundp.ac.be

## ABSTRACT

NIMASTEP is a dedicated numerical software developed by us, which allows one to integrate the osculating motion (using cartesian coordinates) in a Newtonian approach of an object considered as a point-mass orbiting a homogeneous central body that rotates with a constant rate around its axis of smallest inertia. The code can be applied to objects such as particles, artificial or natural satellites or space debris. The central body can be either any terrestrial planet of the solar system, any dwarf-planet, or even an asteroid. In addition, very many perturbations can be taken into account, such as the combined third-body attraction of the Sun, the Moon, or the planets, the direct solar radiation pressure (with the central body shadow), the non-homogeneous gravitational field caused by the non-sphericity of the central body, and even some thrust forces. The simulations were performed using different integration algorithms. Two additional tools were integrated in the software package; the indicator of chaos MEGNO and the frequency analysis NAFF. NIMASTEP is designed in a flexible modular style and allows one to (de)select very many options without compromising the performance. It also allows one to easily add other possibilities of use. The code has been validated through several tests such as comparisons with numerical integrations made with other softwares or with semi-analytical and analytical studies.

The various possibilities of NIMASTEP are described and explained and some tests of astrophysical interest are presented. At present, the code is proprietary but it will be released for use by the community in the near future. Information for contacting its authors and (in the near future) for obtaining the software are available on the web site http://www.fundp.ac.be/en/research/projects/page_view/10278201/

**Key words.** Methods: numerical – Gravitation – Celestial mechanics – Space vehicles

## 1. Introduction

The orbital motion of an artificial satellite, a space debris,or a natural satellite have important similarities: their dynamics mainly depends on the physical parameters of the central body (mass, radius, shape, etc.), the physical parameters of the object under study (mass, shape, reflectivity, etc.) and the dynamics of other bodies orbiting the same central body (for example the influence of the Moon or of the Sun on an Earth's satellite). Therefore, if the initial orbital conditions of the considered body (semi-major axis, eccentricity, inclination, etc.) and the parameters mentioned above are known, the computation of the orbital dynamics can be performed, at a certain level of accuracy, using the same program whatever the exact nature of the problem.

Of course, the orders of magnitude of the forces acting on the body will differ from case to case. For example, the direct solar radiation pressure will have a significant effect on an artificial satellite of the Earth but none on a natural satellite of an asteroid. In contrast, the effect of the non-sphericity of the central body will have a much greater influence on a satellite of asteroid (with a shape usually far from a sphere) than on an artificial satellite of the Earth. Therefore an incomplete dynamical model with forces switched on/off depending to magnitude can be useful to explain the main part of the real motion or to validate an analytical theory (in which we keep only the main effects).

These considerations have led us to create the software NIMASTEP (**N**umerical **I**ntegration of the **M**otion of **A**rtificial **S**atellites orbiting a **TE**lluric **P**lanet). NIMASTEP has been (first) implemented during the Ph.D. thesis of N. Delsate (Delsate 2011b). NIMASTEP was initially developed to numerically integrate the motion of an artificial satellite orbiting a telluric planet (Delsate et al. 2010; Lemaître et al. 2009; Valk et al. 2009a)[1]. Later on this software was extended to calculate the motion of one or several artifical (Delsate 2011a) or natural satellites (Compère et al. 2012) orbiting a (dwarf-)planet or an asteroid of the solar system. In the current version of the software, several forces are still neglected. This will be discussed briefly in section 2.3. But the current state of the code is not the limit, future versions could include these forces step by step. The aim of this software is not to become an orbit restitution software but to be used to analyze the different orbital dynamics. So far, the code can handle quite a few different applications. The first step is always a two-body problem, where the central body is considered as a homogeneous mass and rotates (at constant rate) around its axis of smallest inertia. The small body is distant and small enough to have no effect on the motion of the central body and spherical enough to be considered a point-mass satellite. As we will show below, different perturbations already available can be applied to this initial system.

---

[1] The name of this software has been fixed at this time.

The first characteristic of our software is its extensibility and its generality (within its possibilities): it allows a huge choice of central bodies, of forces, of objects of interest, etc. and can therefore be used by various members of the dynamical or geodesian community. The second originality of our software is the availability of two efficient tools of analysis: the indicator of chaos MEGNO (Cincotta et al. 2003; Cincotta & Simó 2000; Goździewski et al. 2001) and the frequency analysis NAFF (Laskar 1999, 2003, 2005) are directly connected to the numerical integrations.

The code is written in FORTRAN 90 and is divided into modules that can be used in other programs (e.g., the module of the numerical integrators). Each force (gravity field of the central body, radiation pressure, gravitational attraction of a third body, etc.) can be switched on or off. The software has been validated through several methods: comparison with other integrators, checkings with analytical and semi-analytical existing methods and systematic reproduction of known behaviors.

The paper is structured as follows: in section 2, we describe the forces that can be taken into account (or not) by NIMASTEP and we give the related equations of motion. In section 3, we present the two supplementary tools (MEGNO and NAFF). In section 4, we demonstrate the reliability of NIMASTEP by reproducing known results; we present some tests to show that NIMASTEP correctly simulates physical processes by comparisons with another software and with semi-analytical results and analytical studies. In the last section, we summarize and highlight the major features of the software and its future extensions.

The version of NIMASTEP presented in this paper is the web version 5.35w.

## 2. Code capabilities

NIMASTEP numerically integrates the osculating equations of motion in a Newtonian approach of a small body considered as a point-mass around a homogeneous and rotating (with a constant rate) main body. The (fixed) reference frame has its origin at the center of mass of the main body and the x-y plane corresponds to the equatorial plane. The location of the prime meridian (sidereal angle) and the direction of the north pole (right ascension and declination) with respect to the International Celestial Reference Frame (ICRF) are extracted from Archinal et al. (2011).

The position of the small body is written in cartesian coordinates $(x, y, z, \dot{x}, \dot{y}, \dot{z}) \equiv (\boldsymbol{r}, \dot{\boldsymbol{r}} = \boldsymbol{v})$ and $r = |\boldsymbol{r}| = \sqrt{x^2 + y^2 + z^2}$. This choice ensures a high stability for the numerical solution by avoiding both eccentricity and inclination singularities. The equations of motion are given by

$$\begin{cases} \dot{\boldsymbol{r}} = \boldsymbol{v} \\ \dot{\boldsymbol{v}} = \ddot{\boldsymbol{r}}_{\text{pot}} + \sum_i \ddot{\boldsymbol{r}}_{3b_i} + \nu \ddot{\boldsymbol{r}}_{\text{rp}} + \ddot{\boldsymbol{r}}_{\text{th}} \;, \end{cases} \quad (1)$$

where $\ddot{\boldsymbol{r}}_{\text{pot}}$, $\ddot{\boldsymbol{r}}_{3b_i}$, $\nu \ddot{\boldsymbol{r}}_{\text{rp}}$ and $\ddot{\boldsymbol{r}}_{\text{th}}$ are the accelerations due to the gravitational potential of the central body, respectively to the gravitational attraction of the third bodies, to the direct solar radiation pressure (with or without umbra) and to the possible thrust acting on the satellite. These accelerations are developed in section 2.2.

The integrations are performed in the fixed reference frame (defined above). However, the computation of some parts of the acceleration, as $\ddot{\boldsymbol{r}}_{\text{pot}}$, is made in a uniformly rotating frame (of velocity $\dot{\theta}$ around the axis of smallest inertia of the main body). To convert it to the fixed frame, we use

$$\boldsymbol{r}_{\text{fix}} = R_\theta \, \boldsymbol{r}_{\text{rot}} \;, \quad (2)$$

with $\boldsymbol{r}_{\text{fix}}$ ($\boldsymbol{r}_{\text{rot}}$) the location of the satellite in the fixed reference frame (in the rotating reference frame), $\theta$ the sidereal angle (time-dependent) and $R_\theta$ the rotation matrix given by

$$R_\theta = \begin{pmatrix} \cos\theta & -\sin\theta & 0 \\ \sin\theta & \cos\theta & 0 \\ 0 & 0 & 1 \end{pmatrix} . \quad (3)$$

Equation (2) corresponds to the following expression for the acceleration:

$$\ddot{\boldsymbol{r}}_{\text{fix}} = \ddot{R}_\theta \, \boldsymbol{r}_{\text{rot}} + 2 \, \dot{R}_\theta \, \dot{\boldsymbol{r}}_{\text{rot}} + R_\theta \, \ddot{\boldsymbol{r}}_{\text{rot}} \;, \quad (4)$$

where the values of $\theta$ and $\dot{\theta}$ come from Archinal et al. (2011). We consider $\ddot{\theta}$ to be equal to 0.

### 2.1. Numerical integration

The software enables the user to choose the integrator. The present propositions are on the one hand three single-step Runge-Kutta methods (hereafter RK): a Runge-Kutta of order 4 (Hairer et al. 1993) (RK4), a Runge-Kutta-Fehlberg of order 4 with a variable step size (Hairer et al. 1993) (RKF4), and an explicit Runge-Kutta method of order 8(5,3) (DOP853) according to Dormand & Prince (Hairer et al. 1993); on the other hand, some other methods: a Bulirsch-Stoer (Stoer & Bulirsch 1980) with a variable step size, and the multistep method of Adams-Bashforth-Moulton of order 10 (Hairer et al. 1993) (ABM10). Each of these integrators has been tested and validated.

The first method (RK4) is one of the most widely used methods. It is efficient for simple problems with smooth curves. But it is not adapted to problems with high variation of the derivatives, such as highly eccentric orbits. The RKF4, DOP853, and Bulirsch-Stoer use an adaptive step size to improve, for example, the integrations of these orbits. Runge-Kutta method of order 8 (DOP853) and Bulirsch-Stoer are algorithms often used in spatial geodesy and n-body problem integrations. This is why we introduced these integrators in our software. Finally the ABM10 was implemented to obtain a precise constant-step-size integrator faster than RK methods.

The advantage of multistep over single-step RK methods of the same accuracy is that the multistep methods require only one function evaluation per step, while, e.g., the RK4 method requires four, and the RKF4 method, six function evaluations. Conversely, the disadvantage of the multistep methods is the difficulty to change the step size during the integration, while for the single-step RK methods this is an easier procedure. The main disadvantage of explicit methods (like RK) is that they are only conditionally stable, and the critical time step may be several orders of magnitude smaller that the desired time step. The ABM10 overcomes this disadvantage by using first an explicit predictor and second an implicit corrector (unconditionally stable). The advantage of Bulirsch-Stoer is that it is often possible to choose extremely long step sizes while maintaining reasonable accuracy. But the algorithm is more complicated and makes a single step rather slow.

### 2.2. Accelerations

Let us describe the different accelerations (or forces) taken into account.

#### 2.2.1. Non-sphericity of the central body: spherical harmonics expansion

The calculation of the acceleration through the gravity field of the central body is performed by using a spherical harmonics development for the potential (Kaula 1966)

$$U(r, \lambda, \phi) = -\frac{\mu}{r} \sum_{n=0}^{\infty} \sum_{m=0}^{n} \left(\frac{R_e}{r}\right)^n P_{n,m}(\sin \phi) \times \quad (5)$$
$$\left(C_{n,m} \cos(m\lambda) + S_{n,m} \sin(m\lambda)\right),$$

where $\mu = GM$ is the gravitational constant of the central body, $R_e$ is the radius of the main body, $(r, \lambda, \phi)$ are the spherical coordinates of the small body, $C_{n,m}$ and $S_{n,m}$ are physical constants depending on the shape of the main body and named coefficients of the expansion ($n$ is the degree and $m$ is the order of the coefficients) and $P_{n,m}$ are the associated Legendre polynomials.

To calculate the partial derivatives of the potential $U = \mu V$ with respect to the cartesian coordinates, and then the acceleration, we use the accurate and efficient formulation (in the rotating frame) of Cunningham (1970):

$$\ddot{\boldsymbol{r}}_{rot} = \mu \left(\frac{\partial V}{\partial x}, \frac{\partial V}{\partial y}, \frac{\partial V}{\partial z}\right)^T, \quad (6)$$

$$\text{where} \quad \frac{\partial V}{\partial \bullet} = \Re \left\{\sum_{n=0}^{\infty} \sum_{m=0}^{n} R_e \left(C_{n,m} - \imath S_{n,m}\right) \frac{\partial V_{n,m}}{\partial \bullet}\right\},$$

with $\Re\{z\}$ giving the real part of the complex $z$. The intermediate functions $V_{n,m}$ are calculated with the recurrent formulas

$$V_{0,0} = 1/r, \quad (7)$$

$$V_{n,n} = (2n-1) \frac{x + \imath y}{r^2} V_{n-1,n-1}, \quad (8)$$

$$V_{n+1,n} = (2n+1) \frac{z}{r^2} V_{n,n}, \quad (9)$$

$$(n-m)V_{n,m} = (2n-1) \frac{z}{r^2} V_{n-1,m} - \frac{n+m-1}{r^2} V_{n-2,m}, (10)$$

where the coordinates $(x, y, z)$ fix the position of the body in the rotating reference frame. In equation (6), the $\partial/\partial\bullet$ means "partial derivative with respect to a coordinate $x$, $y$ or $z$". These derivatives (first and second) are given by ($\alpha, \beta, \gamma \in \mathbb{N} \mid \alpha + \beta + \gamma \leq 2$)

$$\frac{\partial^{\alpha+\beta+\gamma}}{\partial^\alpha x\, \partial^\beta y\, \partial^\gamma z} V_{n,m} = \imath^\beta \sum_{j=0}^{\alpha+\beta} \frac{(-1)^{\alpha+\gamma-j}}{2^{\alpha+\beta}} \frac{(n-m+\gamma+2j)!}{(n-m)!}$$
$$\times C_{\alpha,\beta,j}\ V_{n+\alpha+\beta+\gamma, m+\alpha+\beta-2j}, \quad (11)$$

where

$$C_{\alpha,\beta,j} = \sum_k (-1)^k \binom{\alpha}{j-k} \binom{\beta}{k}, \quad (12)$$

with $\max\{0, j - \alpha\} \leq k \leq \min\{\beta, j\}$.

Equation (4) is used to calculate the acceleration in the fixed reference frame. The coefficients $C_{n,m}$ and $S_{n,m}$, which depend on the shape of the body, are provided by the user.

#### 2.2.2. Non-sphericity of the central body: MacMillan potential for an ellipsoid

If the shape of the central body is approximated by a tri-axial ellipsoid (with semi-axes $a \geq b \geq c$, $a$ on the x axis, $b$ on the y axis and $c$ on the z axis), the potential induced by this body in the rotating frame is given by the MacMillan formula (MacMillan 1958), reformulated by Garmier & Barriot (2001):

$$V(x,y,z) = -\frac{3\mu}{2} \int_{\lambda_1}^{+\infty} \left(1 - \frac{x^2}{s^2} - \frac{y^2}{s^2 - h^2} - \frac{z^2}{s^2 - k^2}\right) \times$$
$$\frac{ds}{\sqrt{(s^2 - h^2)(s^2 - k^2)}}, \quad (13)$$

where $h^2 = a^2 - b^2 \leq a^2 - c^2 = k^2$. $\lambda_1$ is the first elliptical coordinate of the point $(x, y, z)$ calculated by taking the square root of the largest solution in $s^2$ of the equation

$$\frac{x^2}{s^2} + \frac{y^2}{s^2 - h^2} + \frac{z^2}{s^2 - k^2} = 1. \quad (14)$$

The acceleration caused by this potential has been introduced in NIMASTEP (Compère et al. 2012):

$$\ddot{\boldsymbol{r}}_{rot} = \begin{pmatrix} -\dfrac{3}{2}\dfrac{x\mu}{\lambda_1} \displaystyle\int_{-1}^{1} \dfrac{(t+1)^2}{\alpha^{1/2}(t)\beta^{1/2}(t)} dt \\[2ex] -6\,y\,\mu\,\lambda_1 \displaystyle\int_{-1}^{1} \dfrac{(t+1)^2}{\alpha^{3/2}(t)\beta^{1/2}(t)} dt \\[2ex] -6\,z\,\mu\,\lambda_1 \displaystyle\int_{-1}^{1} \dfrac{(t+1)^2}{\alpha^{1/2}(t)\beta^{3/2}(t)} dt \end{pmatrix}, \quad (15)$$

with $\alpha = 4\lambda_1^2 - h^2\,(t+1)^2$ and $\beta = 4\lambda_1^2 - k^2\,(t+1)^2$. The integrals in (15) are computed with a Gauss-Legendre quadrature.

To obtain the acceleration in the fixed reference frame we again use equation (4).

#### 2.2.3. Third body attraction

The acceleration of a satellite orbiting a central point-mass perturbed by a third body (in a fixed reference frame linked to the central body) is given, for example in Montenbruck & Gill (2000) or Murray & Dermott (1999), by

$$\ddot{\boldsymbol{r}}_{3b} = -G\,\frac{M+m}{\|\boldsymbol{r}\|^3}\,\boldsymbol{r} + G\,M_{3b} \left(\underbrace{\frac{\boldsymbol{r}_{3b} - \boldsymbol{r}}{\|\boldsymbol{r}_{3b} - \boldsymbol{r}\|^3}}_{\text{Direct part}} - \underbrace{\frac{\boldsymbol{r}_{3b}}{\|\boldsymbol{r}_{3b}\|^3}}_{\text{Indirect part}}\right), \quad (16)$$

where $M$, $M_{3b}$, and $m$ are the masses of the central body, of the third body and of the satellite. The position vector $\boldsymbol{r}_{3b}$ of the third body with respect to the central body is given by ephemeris. These ephemeris are either highly accurate, like the JPL DE405 (Standish 1998) and INPOP ephemeris (Fienga et al. 2008; Manche 2010), or correspond to a simple Keplerian motion of the third body.

### 2.2.4. Direct solar radiation pressure

We assume that the small body or the satellite is spherical. The albedo of the central body is ignored and, in a first step, the central body - shadowing effects are not taken into account. The acceleration induced by the direct solar radiation pressure is given by (Montenbruck & Gill 2000)

$$\ddot{\boldsymbol{r}}_{\mathrm{rp}} = P_\odot \, C_r \, \left( \frac{a_\odot}{\|\boldsymbol{r} - \boldsymbol{r}_\odot\|} \right)^2 \frac{A}{m} \frac{\boldsymbol{r} - \boldsymbol{r}_\odot}{\|\boldsymbol{r} - \boldsymbol{r}_\odot\|}, \qquad (17)$$

where $C_r$ is the adimensionnal reflectivity coefficient that depends on the optical properties of the satellite surface; $P_\odot = 4.56 \times 10^{-6}$ N/m$^2$ is the radiation pressure for an object located at the distance of 1 AU; $a_\odot$ is a constant equal to the mean distance between the Sun and the central body, and $r_\odot$ is the planetocentric position of the Sun. Finally, the coefficient $A/m$ is the area-to-mass ratio where $A$ is the effective cross-section and $m$ the mass of the satellite. This coefficient is present also in the (neglected) atmospheric drag effect.

The central body - shadowing effects can be taken into account by multiplying formula (17) by a coefficient $\nu$. In this software, we offer two possibilities to calculate this coefficient. In both cases we assume a spherical central body and we neglect its atmosphere.

First, we assume a cylindric umbra. In this case, the Sun is assumed to be distant enough from the central body so that the solar rays are considered parallel. Two conditions (Montenbruck & Gill 2000) have to be respected to consider that the satellite is in the umbra ($\nu = 0$):

$$\frac{\boldsymbol{r}_\odot \cdot \boldsymbol{r}}{\|\boldsymbol{r}_\odot\| \, \|\boldsymbol{r}\|} < 0 \quad \text{and} \quad \sin \psi \, \|\boldsymbol{r}\| < R_e, \qquad (18)$$

where $\psi$ is the angle between the vector central body to satellite ($\boldsymbol{r}$) and the vector central body to Sun ($\boldsymbol{r}_\odot$). The symbol $\cdot$ designates the scalar product. Out of the shadow the coefficient $\nu$ is equal to 1.

Second, we can take into account the penumbra transition (conical shadow). We use the formula given by Montenbruck & Gill (2000) to calculate the coefficient $\nu$. The satellite is

– completely out of the umbra and penumbra ($\nu = 1$) if $a + b \le c$;
– completely in the umbra ($\nu = 0$) if $c < b - a$ or $c < a - b$;
– in the penumbra otherwise.

In this last case, the coefficient $\nu$ is calculated with

$$\nu = 1 - \frac{a^2 \arccos\left(\frac{x}{a}\right) + b^2 \arccos\left(\frac{c-x}{b}\right) - c \sqrt{a^2 - x^2}}{\pi a^2}, \qquad (19)$$

where

$$x = \frac{c^2 + a^2 - b^2}{2\,c}, \qquad (20)$$

$$a = \arcsin\left(\frac{R_\odot}{\|\boldsymbol{r}_\odot - \boldsymbol{r}\|}\right), \qquad (21)$$

$$b = \arcsin\left(\frac{R_e}{r}\right), \qquad (22)$$

$$c = \arccos\left(\frac{-\boldsymbol{r} \cdot (\boldsymbol{r}_\odot - \boldsymbol{r})}{r \|\boldsymbol{r}_\odot - \boldsymbol{r}\|}\right), \qquad (23)$$

with $R_\odot$ the equatorial radius of the Sun.

### 2.2.5. Thrust acceleration

Assuming that the small body is an artificial satellite, we also include a constant thrust in three directions in the accelerations. We consider the mass of the satellite to be constant. The three allowed directions are the radial direction (subscript 1), the direction of the angular momentum (subscript 2), and the transverse direction (subscript 3). The acceleration is given by (Montenbruck & Gill 2000)

$$\ddot{\boldsymbol{r}}_{\mathrm{th}} = \frac{1}{m} \begin{pmatrix} P_1 \dfrac{\boldsymbol{r}}{\|\boldsymbol{r}\|} \\[1.5ex] P_2 \dfrac{\boldsymbol{r} \times \dot{\boldsymbol{r}}}{\|\boldsymbol{r} \times \dot{\boldsymbol{r}}\|} \times \dfrac{\boldsymbol{r}}{\|\boldsymbol{r}\|} \\[1.5ex] P_3 \dfrac{\boldsymbol{r} \times \dot{\boldsymbol{r}}}{\|\boldsymbol{r} \times \dot{\boldsymbol{r}}\|} \end{pmatrix}, \qquad (24)$$

where $P_i$ is the thrust (kg m/s$^2$) coordinate in the direction $i$, for $i = 1, 2, 3$ and the symbol $\times$ designates the vector product. This option has been implemented and successfully used in Delsate (2011a).

### 2.3. Neglected forces and effects

At the conception of NIMASTEP, we have chosen to introduce forces by decreasing order of magnitude (for our test cases). Therefore some forces are not taken into account in NIMASTEP at the moment, but they could be inserted in next versions of this software.

For example, the general relativity is neglected for two reasons. On the one hand, the allowed choice of the central body is restricted to the planets and asteroids of the solar system, for which the effect of the general relativity is weak. For instance, at the altitude ($a \approx 12\,000$ km) of the LAGEOS satellite, the ratio between the Schwarzschild correction and the attraction of the Earth is close to $1 \times 10^{-9}$. This induces a secular effect on the pericenter, of this satellite, reaching 1 km after five years (Deleflie 2002). On the other hand, if this force is taken into account, we have to include some other forces at the same order of magnitude (see Figure 1).

The atmospheric drag has also been neglected because, except for the Earth, either the models of atmosphere are not available or the central body has no atmosphere. Because our first interest was in high altitude satellite, we neglected this effect in the code's first conception. This is also the reason we did not take the tidal effect into account.

Excluding the atmospheric drag (neglected effect) and the radiation pressure, the shape of an artificial satellite has an insignificant effect on its orbit (Celletti & Sidorenko 2008). For the radiation pressure, we considered a mean area-to-mass ratio. The same choice was made in the STELA software (see section 4.2.1): the CNES (The French Space Agency) decided to use an "equivalent" constant $A/m$ for the dynamics of non-spherical satellites. This is why the spacecraft is assumed to be spherical.

Figure 1 presents graphically the order of magnitude (radial component) of different forces (see Capderou 2005 and Montenbruck & Gill 2000 for simplified formulae). This graphic gives an idea of which forces are important or not with respect to the central body and the altitude of the satellite. This plot teaches us that to take into account the relativistic effect on an artificial satellite of the Earth, we cannot neglected the tidal effect or, for low altitude, the atmospheric drag effect. For a probe orbiting Vesta, the relativistic effect is less than $1 \times 10^{-15}$ km/s$^2$ and the effects of Jupiter or Sun are the strongest.

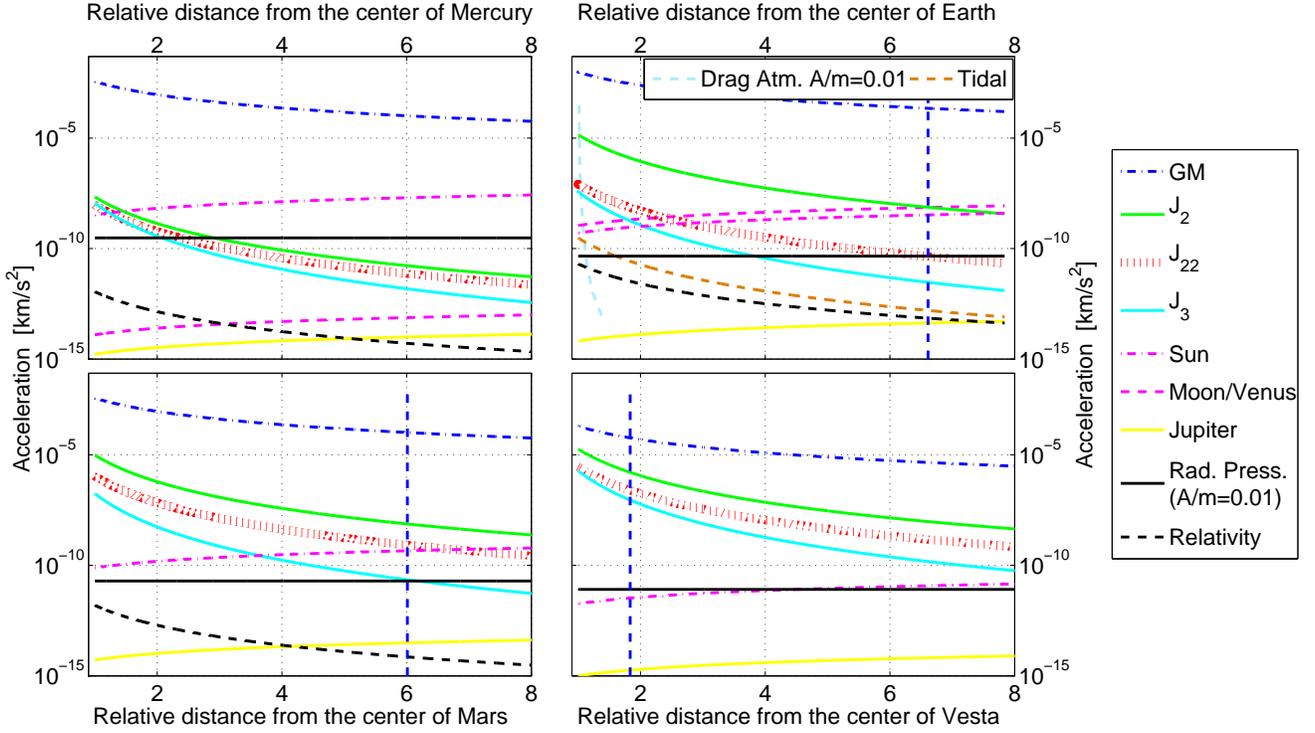

**Fig. 1.** Order of magnitude (radial component) of the main forces acting on a small body orbiting Mercury, Earth, Mars, and Vesta with respect to the relative distance (ratio between the distance from the center of the central body and its equatorial radius) from the center of the central body. The dashed vertical lines localize the synchronous orbits.

## 3. Additional tools

### 3.1. MEGNO: Mean exponential growth factor of nearby orbits

To study the regular or chaotic behavior of the computed orbits, we incorporated an additional tool into NIMASTEP: the chaos indicator MEGNO, which stands for mean exponential growth of nearby orbits (Cincotta & Simó 2000). Here we introduce it briefly.

Let us consider the dynamical system

$$\frac{d}{dt}\boldsymbol{x}(t) = \boldsymbol{f}\big(\boldsymbol{x}(t)\big), \text{ with } \boldsymbol{x} \in \mathbb{R}^{2n}. \quad (25)$$

Let $\phi(t)$ be a solution of the flow defined by equation (25), then the evolution of a tangent vector $\boldsymbol{\delta}(t)$ along $\phi(t)$ is given by

$$\dot{\boldsymbol{\delta}} = \boldsymbol{J}\big(\phi(t)\big) \, \boldsymbol{\delta}(t), \quad \text{with} \quad \boldsymbol{J}\big(\phi(t)\big) = \frac{\partial \boldsymbol{f}}{\partial \boldsymbol{x}}\big(\phi(t)\big). \quad (26)$$

$\boldsymbol{J}$ is the Jacobian matrix of the system (25). The MEGNO indicator, based on a modified time-weighted version of the integral form of the Lyapunov characteristic number, is defined by

$$Y(t) = \frac{2}{t} \int_0^t \frac{\dot{\boldsymbol{\delta}} \cdot \boldsymbol{\delta}}{\|\boldsymbol{\delta}\|^2} s \, ds, \quad (27)$$

and the mean MEGNO is

$$\overline{Y}(t) = \frac{1}{t} \int_0^t Y(s) \, ds. \quad (28)$$

This quantity is a characterization of the divergence rate of two nearby orbits. For stable periodic orbits, $\overline{Y}$ converges toward 0, for stable quasi-periodic orbits, $\overline{Y}$ converges toward 2 and for chaotic motion, $\overline{Y}$ tends toward infinity (Cincotta et al. 2003). In NIMASTEP, we adopt the same strategy as Goździewski et al. (2001) for the numerical computation of the MEGNO.

This indicator has been successfully used in some papers, for instance, Breiter et al. (2005), Cincotta & Simó (2000), Goździewski et al. (2001, 2008) and Hinse et al. (2010, 2008). For the results concerning the MEGNO implemented into NIMASTEP we refer the reader to Valk et al. (2009a) and Compère et al. (2012).

To integrate the system (26), we need the expression of the Jacobian matrix $\boldsymbol{J}$. In NIMASTEP, the MEGNO indicator can be calculated for motions including the nonsphericity of the central body's gravity field, the third-body effect and the direct radiation pressure without shadow ($\nu = 1$). For these three contributions, the Jacobian matrix has the same structure:

$$\boldsymbol{J} = \begin{pmatrix} 0 & 0 & 0 & 1 & 0 & 0 \\ 0 & 0 & 0 & 0 & 1 & 0 \\ 0 & 0 & 0 & 0 & 0 & 1 \\ \left(\boldsymbol{J}^{\text{part.}} \equiv \frac{\partial \ddot{\boldsymbol{r}}}{\partial \boldsymbol{r}}\right) & 0 & 0 & 0 \\ & & & 0 & 0 & 0 \\ & & & 0 & 0 & 0 \end{pmatrix}. \quad (29)$$

Because these forces are conservative, the (partial Jacobian) matrix $\boldsymbol{J}^{\text{part.}}$ is symmetric and its trace is equal to 0. Let us give the $\boldsymbol{J}^{\text{part.}}$ for the three cases.

For the non-sphericity of the central body, given by the spherical harmonic representation, the Cunningham (1970) method gives

$$
\boldsymbol{J}_T^{\text{part.}} = \mu \begin{pmatrix} \dfrac{\partial^2 V}{\partial x^2} & \dfrac{\partial^2 V}{\partial x \partial y} & \dfrac{\partial^2 V}{\partial x \partial z} \\[2mm] \dfrac{\partial^2 V}{\partial y \partial x} & \dfrac{\partial^2 V}{\partial y^2} & \dfrac{\partial^2 V}{\partial y \partial z} \\[2mm] \dfrac{\partial^2 V}{\partial z \partial x} & \dfrac{\partial^2 V}{\partial z \partial y} & \dfrac{\partial^2 V}{\partial z^2} \end{pmatrix}, \quad (30)
$$

with $\dfrac{\partial^2 V}{\partial \bullet \partial \star} = \Re \left\{ \sum_{n=0}^{\infty} \sum_{m=0}^{n} R_e \left( C_{n,m} - i S_{n,m} \right) \dfrac{\partial^2 V_{n,m}}{\partial \bullet \partial \star} \right\}.$ (31)

The second derivatives are given by (11) and this Jacobian matrix (30) is calculated in the rotating reference frame.

The Jacobian matrix due to the non-uniformity of the central body is given by

$$
\boldsymbol{J}_{\text{pot}}^{\text{part.}} = R_\theta^{-1} \, \boldsymbol{J}_T^{\text{part.}} \, R_\theta = R_\theta^T \, \boldsymbol{J}_T^{\text{part.}} \, R_\theta \,. \quad (32)
$$

If the central body is approximated by an ellipsoid, we also use equation (32) where $\boldsymbol{J}_T^{\text{part.}}$ is calculated by (Compère et al. 2012)

$$
\boldsymbol{J}_T^{\text{part.}} = \begin{pmatrix} \dfrac{\ddot{x}_{rot}}{x} + f_1 \dfrac{\partial \lambda_1}{\partial x} & f_1 \dfrac{\partial \lambda_1}{\partial y} & f_1 \dfrac{\partial \lambda_1}{\partial z} \\[2mm] f_2 \dfrac{\partial \lambda_1}{\partial x} & \dfrac{\ddot{y}_{rot}}{y} + f_2 \dfrac{\partial \lambda_1}{\partial y} & f_2 \dfrac{\partial \lambda_1}{\partial z} \\[2mm] f_3 \dfrac{\partial \lambda_1}{\partial x} & f_3 \dfrac{\partial \lambda_1}{\partial y} & \dfrac{\ddot{z}_{rot}}{z} + f_3 \dfrac{\partial \lambda_1}{\partial z} \end{pmatrix}, \quad (33)
$$

with $f_1 = f_1(\lambda_1, h, k, x) = \dfrac{3x\mu}{\lambda_1^2 (\lambda_1^2 - h^2)^{1/2} (\lambda_1^2 - k^2)^{1/2}},$

$\quad f_2 = f_2(\lambda_1, h, k, y) = \dfrac{3y\mu}{(\lambda_1^2 - h^2)^{3/2} (\lambda_1^2 - k^2)^{1/2}},$ (34)

$\quad f_3 = f_3(\lambda_1, h, k, z) = \dfrac{3z\mu}{(\lambda_1^2 - h^2)^{1/2} (\lambda_1^2 - k^2)^{3/2}}.$

$\ddot{x}_{rot}, \ddot{y}_{rot}$ and $\ddot{z}_{rot}$ are the three components of the acceleration (15) and the partial derivatives of $\lambda_1$ are given in the appendix of Compère et al. (2012).

The partial Jacobian matrix of the third-body effect is given by (Montenbruck & Gill 2000)

$$
\boldsymbol{J}_{3b}^{\text{part.}} = -\mu_{3b} \left( -\dfrac{3}{\|\boldsymbol{r} - \boldsymbol{r}_{3b}\|^5} (\boldsymbol{r} - \boldsymbol{r}_{3b}) \otimes (\boldsymbol{r} - \boldsymbol{r}_{3b})^T \right.
$$
$$
\left. + \dfrac{1}{\|\boldsymbol{r} - \boldsymbol{r}_{3b}\|^3} \boldsymbol{I}_{3\times3} \right), \quad (35)
$$

where $\otimes$ is the tensorial product and $\boldsymbol{I}_{3\times3}$ is the unitary matrix of dimension 3.

Last, for the direct radiation pressure (for $\nu = 1$), the partial Jacobian matrix is similarly calculated using (Montenbruck & Gill 2000)

$$
\boldsymbol{J}_{rp}^{\text{part.}} = P_\odot \, C_r \, a_\odot^2 \, \dfrac{A}{m} \left( -\dfrac{3}{\|\boldsymbol{r} - \boldsymbol{r}_\odot\|^5} (\boldsymbol{r} - \boldsymbol{r}_\odot) \otimes (\boldsymbol{r} - \boldsymbol{r}_\odot)^T \right.
$$
$$
\left. + \dfrac{1}{\|\boldsymbol{r} - \boldsymbol{r}_\odot\|^3} \boldsymbol{I}_{3\times3} \right). \quad (36)
$$

## 3.2. NAFF: Numerical analysis of the fundamental frequencies

Another tool directly integrated into NIMASTEP is the frequency analysis introduced by J. Laskar (Laskar 1999, 2003, 2005). The NAFF (numerical analysis of the fundamental frequencies) is a very efficient numerical method for a frequency analysis, much more efficient than a classical FFT. Indeed, for a KAM solution, the accuracy of the frequencies of a signal on a time span $[-T, T]$ is proportional to $1/T^4$ for the NAFF, using the Hanning window, while for an ordinary FFT method, this accuracy is only proportional to $1/T$ (Laskar 1999).

The main purpose of the NAFF is to determine the approximation $f^\bullet(t)$ (with a given number N of harmonics) of a signal $f(t)$ from its numerical knowledge, where both functions $f$ and $f^\bullet$ are developed in Fourier series:

$f^\bullet(t) = \sum_{k=1}^{N} a_k^\bullet \, e^{i \nu_k^\bullet t}$ is the approximation of the initial signal $f(t) = \sum_{k=1}^{\infty} a_k \, e^{i \nu_k t}$, where $\nu_k$ and $\nu_k^\bullet$ are the real frequencies, when $a_k$ and $a_k^\bullet$ the complex amplitudes. The frequencies $\nu_k^\bullet$ for $k = 1, \ldots, N$ and their associated decreasing amplitudes $a_k^\bullet$ for $k = 1, \ldots, N$ are determined through an iterative scheme. Indeed, to determine the first frequency $\nu_1^\bullet$, one searches for the maximum of the amplitude of

$$
\phi(\nu) = \langle f(t), \exp(\imath \nu t) \rangle, \quad (37)
$$

where the scalar product $\langle f(t), g(t) \rangle$ is defined by

$$
\langle f(t), g(t) \rangle = \frac{1}{2T} \int_{-T}^{T} f(t) \, \overline{g(t)} \, \chi(t) \, dt, \quad (38)
$$

where $\overline{g(t)}$ is the complex conjugate of $g(t)$ and where $\chi(t)$ is a weight function, i.e., a positive function verifying

$$
\frac{1}{2T} \int_{-T}^{T} \chi(t) \, dt = 1. \quad (39)
$$

Laskar (1999) recommends the following choice (Hanning)

$$
\chi(t) = \frac{2^p (p!)^2}{(2p)!} \left( 1 + \cos(\pi t) \right)^p, \quad (40)
$$

where $p$ is a positive integer. In practice, the algorithm is more efficient when $p = 1$ or $p = 2$.

Once the first periodic term $\exp(\imath \nu_1^\bullet t)$ is found, its complex amplitude $a_1^\bullet$ is obtained by orthogonal projection, and the process is restarted with the remainder $f_1(t) = f(t) - a_1^\bullet \exp(\imath \nu_1^\bullet t)$ as initial input. The algorithm stops when two detected frequencies are too close to each other, because this alters their determinations, or when the number of detected terms reaches a maximum set by the user. When the difference between two frequencies is larger than twice the frequency associated with the length of the total time interval, the determination of each fundamental frequency is not perturbed by the other ones.

To our knowledge, the first uses of NAFF in the geodesy context were published by Delsate & Lemaître (2009), Delsate et al. (2008, 2009) and Valk et al. (2009a). In Valk et al. (2009a), the authors (using NIMASTEP and NAFF) computed the frequency of the resonant angle of the geostationary orbit. Their results agree well with the theory of Valk et al. (2009b).

It is possible to use NAFF for other purposes than the decomposition of a signal into frequencies: it can be used as an indicator of dissipation of the main frequencies and therefore as a chaos indicator. This can be achieved in two ways. The first possibility is to calculate the variations of the main frequency (let us call it $\nu$) of a signal with respect to the initial conditions (here e.g. $x$ and $y$). We use the frequency numerical second derivative, $\delta\delta\nu$, given by the formula (Laskar 1993)

$$\begin{aligned} \delta\delta\nu(x,y) \ = \ & |\nu(x,y) - 2\nu(x - \Delta x, y) + \nu(x - 2\Delta x, y)| \\ & + \ |\nu(x,y) - 2\nu(x, y - \Delta y) + \nu(x, y - 2\Delta y)|. \quad (41) \end{aligned}$$

With $\log(\delta\delta\nu(x,y))$, we can detect the irregularities in a frequency map. The second possibility is to calculate the variations of the main frequency of a signal with respect to the time; we compute a first value $\nu^{(1)}$ of the main frequency of the signal on a time span $T_1$ (e.g. from $-T$ to 0), then we compute a second value $\nu^{(2)}$ of the same signal on a time span $T_2$ (e.g. from 0 to $T$). Consequently, we estimate the diffusion rate with respect to the time by

$$\log\left(\frac{\nu^{(1)} - \nu^{(2)}}{\nu^{(1)}}\right). \quad (42)$$

Because the two indicators (41 and 42) give approximately the same results, they can be used interchangeably (Laskar 1999). The computation of the second-order derivatives usually requires very small stepsizes $\Delta x$ and $\Delta y$, while the computation of the time diffusion requires more iterations.

These two methods were successfully used with NIMASTEP in Lemaître et al. (2009) and Delsate et al. (2010).

# 4. Validation

We performed very many tests to prove that the algorithms programed in NIMASTEP were correctly implemented. Because it is not possible to describe all tests, we chose to show some validations of the code related to applications (already published or not). We present different comparisons with an external full numerical integration, semi-analytical studies, and some known analytic solutions.

We usually do not have an easy access to real data, therefore comparisons with these data are difficult to perform. In addition, it is much easier and rigorous to compare the performance or results with that of other codes or with analytical results. The main reason is that for these data we have all informations on the approximations. Real data are not ideal to validate a code based on an approximate model of reality.

## 4.1. Comparisons with full numerical integrations

For a direct comparison (section 4.1.1), we compare NIMASTEP with "real" data and with a software written by A. Rossi called R-ISTIs in the following. For an indirect comparison (section 4.1.2), we compare measures (as the evolution of the MEGNO or the semi-major axis variation) computed through integrations obtained by NIMASTEP with external published results.

### 4.1.1. Direct comparison

The first validation is the comparison of an integration performed by NIMASTEP with "real" data and with the results of R-ISTIs (*personal communication of A. Rossi*) using the same initial conditions, parameters and, forces.

First, we briefly describe the TLE (two-line elements) data base (because it is used by R-ISTIs and to compare our software with "real" data.) A two-line element[2] is a data format used by NASA, to describe the orbital elements of the orbit of an Earth satellite. The elements in the TLE sets are mean elements calculated to fit a set of observations using a specific model: the SGP4/SDP4 orbital model (Hoots & Roehrich 1980). The mean orbital elements produced by the SGP4/SDP4 orbital model are Earth-centered-inertial coordinates with respect to the true equator and the mean equinox of epoch. They do not include the effect of nutation.

It is important to note that the arithmetic or geometric means of osculating data do not have the same value as the TLE. In the same way, different orbital models give different mean elements. Indeed, the TLE of a satellite are neither the real osculating Keplerian elements, nor the mean ones. However, we consider that the TLE gives a *good idea* of the real orbit.

Second, we present a short description of the method used in R-ISTIs. R-ISTIs converts the TLE of the satellite Etalon-1 (catalog number 19751) to osculating elements using the SGP4 theory (Hoots & Roehrich 1980). Then, the integrations are realized thanks to a special perturbation propagator with an integrator based on Cowell's method (Lyddane & Cohen 1962; Yoon et al. 1997) for the numerical integration of the equations of motion. The force model includes the Earth's gravity potential, the lunisolar gravitational perturbation, the solar radiation pressure with eclipses, and several atmospheric density models for air drag computation. This propagator was assembled in Pisa, at ISTI[3], based on the NASA program called ASAP-artificial satellite analysis program (Kwok 1987) and used, for example, for some of the propagations in Rossi (2008). At the Julian date 2448135.5, the initial conditions of this simulation, calculated by A. Rossi (R-ISTIs), are $a = 25\,501.226\,477$ km, $e = 0.642\,773\,427 \times 10^{-3}$, $i = 1.132\,591\,133$ rad, $\Omega = 2.726\,705\,844$ rad, $\omega = 4.284\,489\,314$ rad and $M = 0.243\,368\,293$ rad.

The model of forces, used in NIMASTEP for this test, includes the Earth's attraction developed in a spherical harmonic expansion up to degree and order 20 (the Earth gravity field adopted is the EGM96 model, Lemoine 1987), as well as the combined attractions of the Sun and the Moon (ephemeris DE405 given by the JPL, Standish 1998). The perturbing effects of direct solar radiation pressure are also taken into account for an area-to-mass ratio fixed at $10^{-3}$ m²/kg with conical shadow. At this altitude, the atmospheric drag is insignificant.

We chose the ABM10 integrator with a step size of 200 seconds of time. Figure 2 (upper panels) shows the evolution of the Keplerian elements for both softwares showed in the $(\sin i/2 \cos\Omega, \sin i/2 \sin\Omega)$ and $(e\cos\omega, e\sin\omega)$ planes. We also overlaid the data from the TLE data base

---

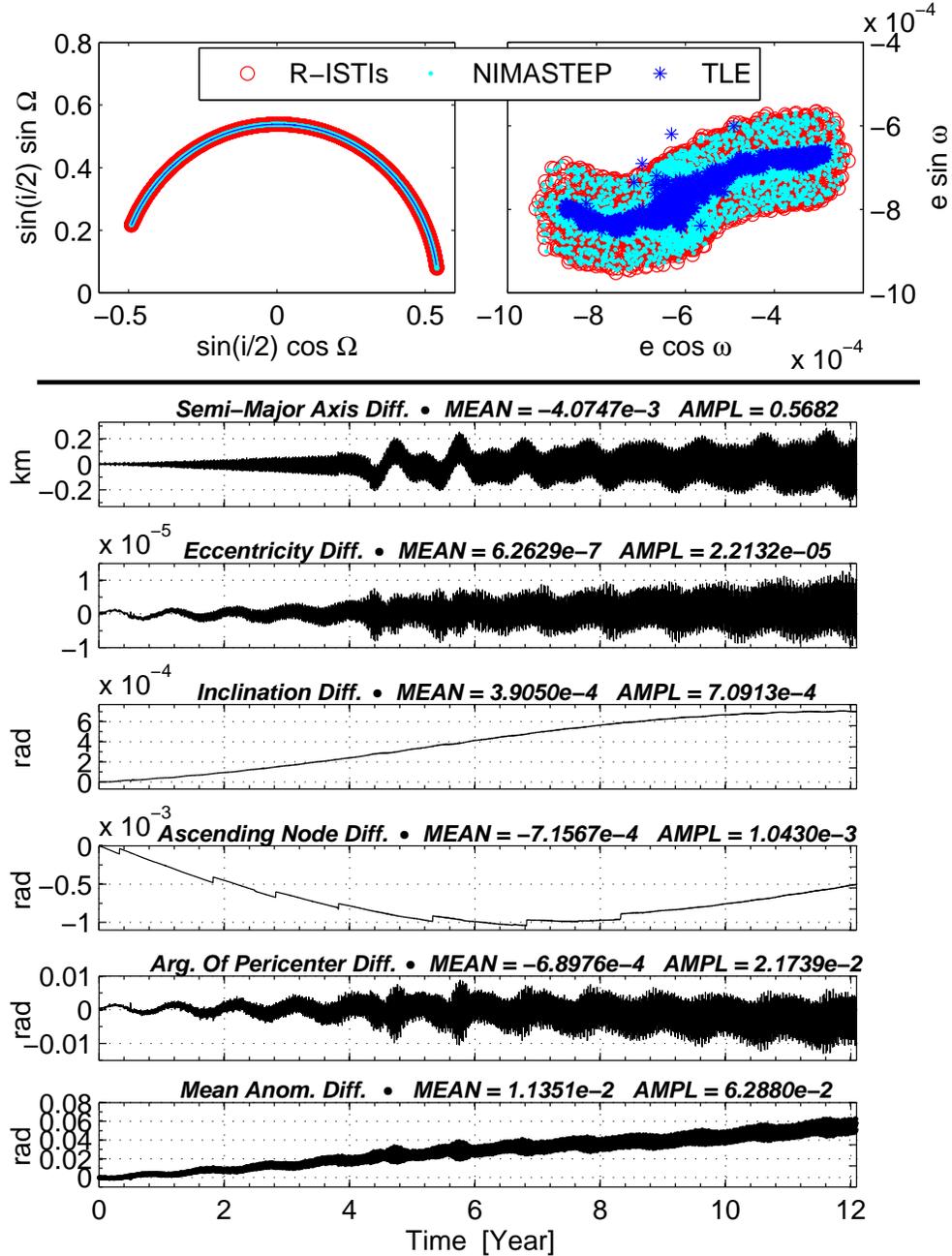

**Fig. 2.** Comparison of NIMASTEP with R-ISTIs and with the TLE data of Etalon-1. Two upper panels: projection of the orbits performed with NIMASTEP (light blue dots), R-ISTIs (red circles), and TLE (blue stars) on the phase spaces $(\sin i/2 \cos \Omega, \sin i/2 \sin \Omega)$ and $(e \cos \omega, e \sin \omega)$. We added $2.7 \times 10^{-4}$ to $e \sin \omega$ of the TLE to obtain a good match. Six lower panels: differences between NIMASTEP and R-ISTIs orbits.

with a $2.7 \times 10^{-4}$ correction added to $e \sin \omega$ to obtain a good agreement. This correction is necessary because the initial conditions taken by R-ISTIs and NIMASTEP are osculating elements and are not exactly the same as the TLE, probably deriving from the definition of the TLE. Remember that TLE are sets of mean elements but not equal to the mean of the osculating Keplerian elements. This is also why we do not present any differences (which will be irrelevant) between TLE and NIMASTEP results. Both softwares give results that agree with the TLE except for the eccentricity and the argument of pericenter, where the TLE data were parallelly shifted up.

There are some differences between the orbit of R-ISTIs and NIMASTEP, but the errors (six lower panels of Figure 2) are very low, which proves the reliability of NIMASTEP in this context. The amplitude of the errors increases with respect to the time, but the mean error stays quasi-constant except for the mean anomaly. This can be explained by a difference of formulation of the Earth rotation between both softwares.

### 4.1.2. Indirect comparison

Tricarico & Sykes (2010) and Delsate (2011a) studied the main gravitational resonances (resonances between the revolution of the satellite and the rotation of the asteroid, hereafter denoted by GR) appearing around the asteroid (4) Vesta. A part of these works consists in calculating the probability of capture of the Dawn spacecraft mission in the 1:1 GR of Vesta.

Tricarico & Sykes (2010) performed this test through a full numerical software. They simulated a slow descent of Dawn from an initial radius of 1 000 km using a 20 mN of thrust and obtained a probability of capture equal to $1/12 = 8.33\%$. They also observed that higher thrusting (until 50 mN) can exhibit a similar behavior. Taking the same initial conditions (semi-major axis) and forces model, Delsate (2011a) performed numerical integrations with 50 000 random mean anomaly with NIMASTEP for 32 different thrusts from 20 mN to 36 mN. He obtained a mean probability of 8.262%, which agrees well (indirectly) with the results obtained by both softwares. That also shows that NIMASTEP is well adapted to this type of study.

### 4.2. Comparisons with semi-analytical theories

As a second validation of our software, we applied NIMASTEP to the typical case of an high-altitude abandoned space debris and compared our results with two semi-analytical theories: the STELA software and a semi-analytical study of Valk et al. (2008, 2009b).

### 4.2.1. STELA software

The semi-analytic tool for end of life analysis (STELA)[4] is a semi-analytical software designed by the CNES, which computed efficient long-term propagations of LEO (low earth orbit), GEO (geostationary earth orbit), and GTO (geostationary transfer orbit) orbits based on semi-analytical models. It allows the computation of "graveyard orbits" and tests the safety of protected zones.

STELA computes the mean orbit parameters by a semi-analytical model (without short periodic motion) at each integration time step. To compute the osculating parameters at a given time, STELA uses the mean (averaged) elements and artificially adds the short periodic effects. The short periodic computation model contains the influence of the oblateness ($C_{20}$) of the Earth, the solar and lunar gravity, and the solar radiation pressure. The mean motion (middle- and long-term evolution of the orbital parameters) includes a complete degree and order 4 model of the gravity field for the Earth, the solar and lunar gravity and the solar radiation pressure. The Earth's shadow is cylindric and the object is assumed to follow a quasi-circular orbit. The integrator is a sixth-order Runge-Kutta method with a step size of one day. More explanations are available, for instance, in Deleflie et al. (2005, 2010), Fraysse (2011), and Le Fevre et al. (2011).

We compared the osculating motion given by NIMASTEP (with the same contributions) with the result of STELA. We chose the following initial conditions: $a = 42\,164.140$ km, $e = 0.001$, $i = 0.1$ rad, and $\Omega = \omega = M = 0$ rad. The initial time at epoch is 2451350.5 JD and the area-to-mass ratio is equal to 0.1 m²/kg with a reflectivity coefficient equal to 1. The integrator is ABM10 with a step size of 864 sec. Figure 3 shows the differences between both integrations (in absolute error). We notice that after 100 years, the maximal error is 1.911 km for the semi-major axis, $2 \times 10^{-4}$ for the eccentricity, 30.94 arcmin for the inclination, and 214.51 arcmin for the ascending node. The amplitude of the error on the argument of pericenter may be caused by a small phase difference in this motion. Another possible explanation is that the method used by both softwares to obtain the pericenter from the cartesian coordinates or from the equinoctial elements are different. Moreover, the eccentricity is very low, inducing an eventual difficulty to determinate the pericenter angle.

The errors are very low for this long time of integration and we conclude that the results NIMASTEP agree well with those of the STELA software.

### 4.2.2. Valk semi-analytical study

The second validation by a semi-analytical study is achieved by comparing our results with the mean motion given by the theory of S. Valk, who studied the geostationary Earth orbits (Valk et al. 2008, 2009b).

Directly comparing the results of the osculating numerical integration (NIMASTEP) with a semi-analytical solution is not entirely consistent. Indeed, the osculating elements and the mean ones are not directly comparable. However, we consider that our main purpose is to give a *first order* comparison.

We start with a short description of the semi-analytical theory; for more details we refer the reader to Valk et al. (2008, 2009b). The framework is a mean motion semi-analytical theory: only the long-term and secular effects are derived. In other words, the resulting theory does not include any short-term effects. In practice, the theory consists of the numerical integration of the filtered equations of motion over the short periods.

First, the approach is to choose a nonsingular and canonical set of variables, namely the Poincaré variables:

$$
\begin{aligned}
x_1 &= \sqrt{2\,(L-G)}\,\sin(-\omega-\Omega)\,, \\
x_2 &= \sqrt{2\,(G-H)}\,\sin(-\Omega)\,, \\
x_5 &= \sqrt{2\,(G-H)}\,\cos(-\Omega)\,, \\
x_3 &= \lambda = \Omega + \omega + M\,, \\
x_4 &= \sqrt{2\,(L-G)}\,\cos(-\omega-\Omega)\,, \\
x_6 &= L\,,
\end{aligned}
\tag{43}
$$

where $L = \sqrt{\mu a}$, $G = L\sqrt{1-e^2}$ and $H = G\cos i$ are the Delaunay elements. Valk et al. (2008, 2009b) expanded the Earth potential in powers of the eccentricity and of the inclination, truncated the development at a low order to obtain an Hamiltonian formulation. Some other perturbations were added, such as the direct radiation pressure (Valk et al. 2008) and the solar and lunar gravitational effects. Second, the final expansion of the Hamiltonian is set in

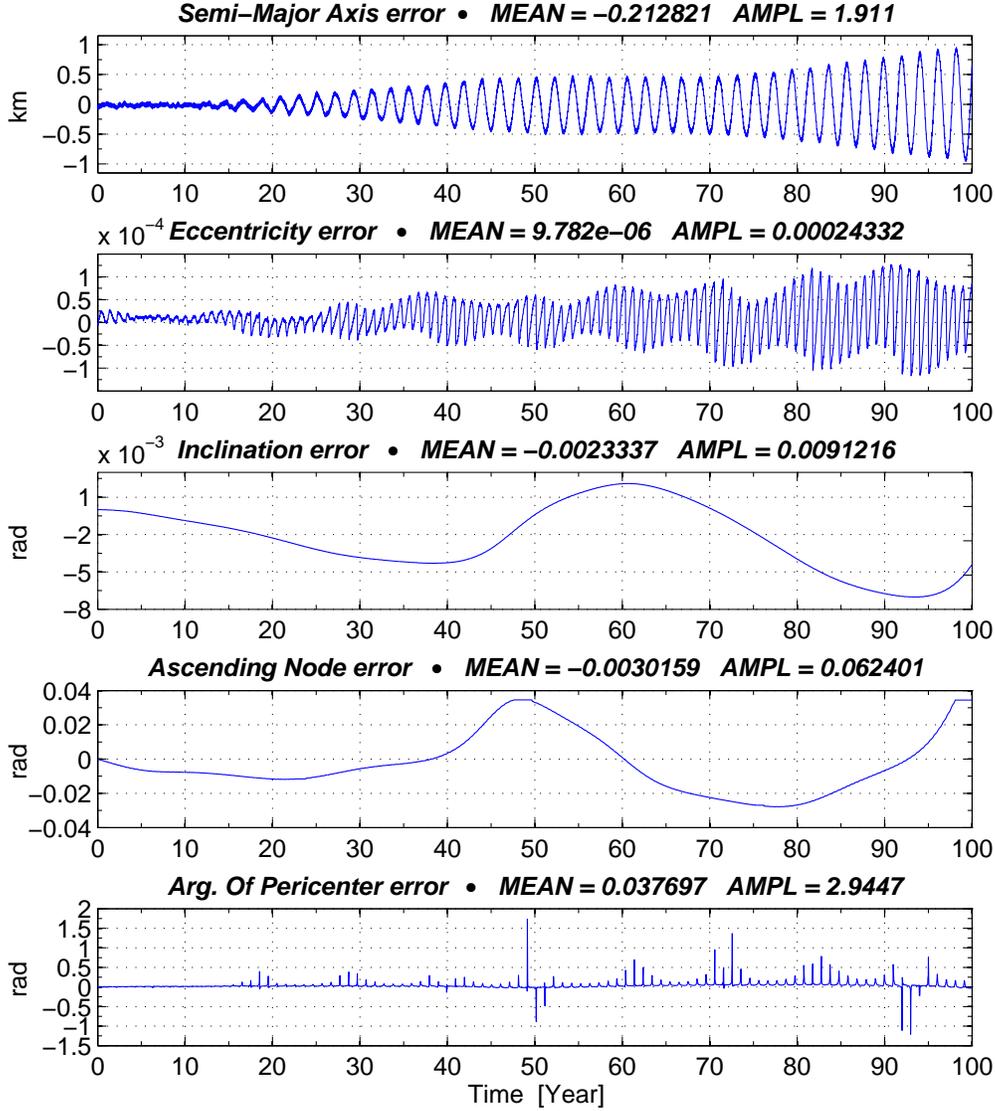

**Fig. 3.** Absolute differences between NIMASTEP and STELA for space debris motions around the Earth. The initial conditions are $a = 42\,164.140$ km, $e = 0.001$, $i = 0.1$ rad and $\Omega = \omega = M = 0$ rad.

non-dimensional cartesian variables using

$$
\begin{aligned}
X_1 &= \sqrt{\tfrac{2\,(L-G)}{L}}\,\sin(-\omega-\Omega)\,,\\
X_2 &= \sqrt{\tfrac{2\,(G-H)}{L}}\,\sin(-\Omega)\,,\\
X_3 &= \lambda\,,\\
Y_1 &= \sqrt{\tfrac{2\,(L-G)}{L}}\,\cos(-\omega-\Omega)\,,\\
Y_2 &= \sqrt{\tfrac{2\,(G-H)}{L}}\,\cos(-\Omega)\,,\\
Y_3 &= L\,.
\end{aligned}
\tag{44}
$$

Last, they averaged the disturbing function over the fast variable, namely the sidereal time $\theta$, to the first order by dropping the fast periodic terms in the trigonometric expansions. The Hamiltonian is written as a *Poisson series* using the symbolic software MSNAM, the series manipulator of Namur (Henrard 1986).

For our validation, let us consider the dynamical evolution of a theoretical high-altitude space debris. The perturbations taken into account are the oblateness ($C_{20}$) of the Earth and the third-body perturbation induced by the Moon. The integration time is equal to 100 years. The entry-level step size used in the osculating numerical integration with NIMASTEP is fixed at 200 seconds (with the RK4 method), whereas Valk et al. (2008) define a one day step size to integrate the averaged system of equations (also with the RK4 method).

To quantitatively estimate the accuracy of our software, Figure 4 shows the differences between both orbits. First, the differences remain small although the chosen time of integration is very sizable. Indeed, the computed root mean square (RMS) can be considered to be very small because the comparison is made without any preliminary fitting of the mean initial conditions (Deprit 1969; Henrard 1970; Valk et al. 2009a). Moreover, the RMS on both the argument of perigee and the longitude of the ascending node are mostly influenced by the singular transformation when projecting the integrated non-singular state vector into Keplerian orbital elements.

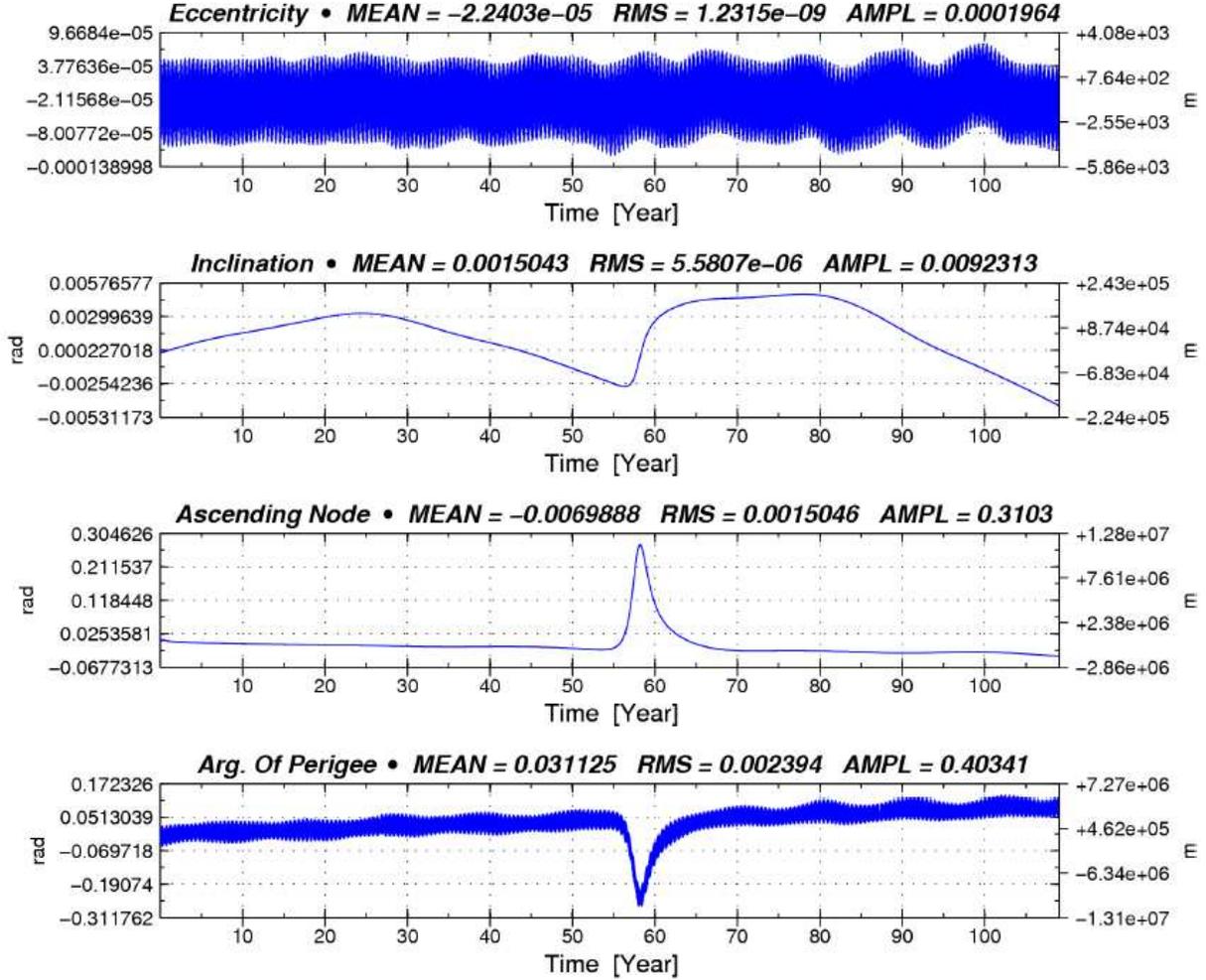

**Fig. 4.** Differences between the osculating orbit and the mean orbit for the long-term time evolution of typical abandoned space debris subject to the oblateness of the Earth ($C_{20}$) and to the Moon perturbation. Initial conditions are $a_0 = 44\,164$ km, $e_0 = 0$, $i_0 = \omega_0 = \lambda_0$ rad. Initial time at epoch is 25 January 1991.

### 4.3. Comparison with results of analytical studies

To validate our software yet again, we compared some results of NIMASTEP with predictions of analytical studies. Some previous articles have shown a good agreement between analytical results and numerical integrations provided by NIMASTEP. In this paper, we only give the context of each analytical study and how our software is validated. The reader is refered to these articles for more details on the results and comparisons.

Valk et al. (2009a,b) and Lemaître et al. (2009) studied the motion of space debris with or without high area-to-mass ratio close to the geostationary Earth orbit (GEO). They take into account the ellipticity of the equator ($C_{22}$) and the direct solar radiation pressure. At first, Valk et al.

(2009a) assumed $A/m = 0$ and localized numerically (with NIMASTEP) the position of the GEO equilibrium. The analytical study of Valk et al. (2009b) yielded the same location. Second, Valk et al. (2009a) performed a frequency analysis (with NAFF in NIMASTEP) of the resonant angle evolution. They obtained a fundamental frequency of libration differing from 0.44% with the analytical estimation (Chao 2005; Valk et al. 2009b; Vallado 2001). Third, increasing the area-to-mass ratio, Valk et al. (2009a) discovered (always helped by NIMASTEP) a web of secondary resonances close to the GEO resonance. Delsate & Lemaître (2009) and Lemaître et al. (2009) analytically explained, located and listed some of the fundamental frequencies.

Delsate et al. (2010) studied the stability of a massless probe orbiting an oblate central body perturbed by a third body. They analytically determined the location of frozen orbits ($\dot{e} = \dot{\omega} = 0$). They also computed the periods of the equilibria. Moreover, for the particular case of a probe around Mercury, they used NIMASTEP to validate their results. They satisfactorily recovered the locations, orbits and periods of the equilibria numerically.

As previously said (section 4.1.2), Delsate (2011a) studied the main GR appearing around the asteroid (4) Vesta. NIMASTEP was used to obtain a map of the radius variation of a spacecraft bringing the GR to the fore. These numerical results agree well with those of Tricarico & Sykes (2010) and with the analytical study.

Compère et al. (2012) studied the evolution of a small body in rotation around a rigid ellipsoidal asteroid in constant rotation. Simulations were made with NIMASTEP using the MacMillan (1958) potential for the influence of the shape of the asteroid on the small body and the chaos indicator MEGNO. The authors presented stability maps showing extremely stable conic-like curves corresponding to GR. An analytical model, based on a simplification of the MacMillan potential, was created to validate the numerical results. Here again, the analytical model confirms the results performed with NIMASTEP.

## 5. Summary and perspectives

The dedicated software NIMASTEP allows one to numerically integrate the osculating motion of an object (artificial or natural satellite, space debris, small natural body) orbiting a telluric planet, a (dwarf-)planet, or an asteroid of the solar system. The equations of motion, in the Newtonian formalism, can include different types of perturbations, such as the influence of one (or more) third body (bodies), of the non-spherical shape of the central body, of the radiation pressure, or even of a thrust acceleration. This software includes two supplementary tools to analyze the orbits: the chaos indicator MEGNO and the frequency analysis NAFF.

It is flexible and modular, which allows very many options (forces, central bodies, object characteristics) to be selected or disabled easily and, in the future, to add some possibilities and the neglected forces. By this extensibility, this software could be used by many members of the dynamical or geodesian community. It has been successfully validated by several astrophysical interest tests, comparisons with other numerical integrations, with semi-analytical results, and finally with analytical studies.

The code is not available yet on the internet, but a web site provides information for contacting its authors and fo obtaining (in the near future) the software: http://www.fundp.ac.be/en/research/projects/page_view/10278201/.

The most important future additions scheduled for NIMASTEP are (I) to include the atmospheric drag of the central body, (II) to simultaneously calculate the motion of more than one (interacting) satellites, (III) to include the general relativity (Will 2006), (IV) to include tidal effects, (V) to include the attitude motion of the satellite, (VI) to modelize the gravity field by other methods, and (VII) to add YORP effects.

<u>Acknowledgements.</u> The authors thank A. Rossi and S. Valk for the simulations provided to validate our software. They also thank A. Lemaitre for her careful reading. The authors thank the referee for her/his constructive comments about the preliminary version of the paper, and B. Noyelles and F. Deleflie for fruitful discussions. The accommodation of NIMASTEP for the web version has been made possible thanks to F. Wautelet.